\documentclass[12pt]{iopart}

 \usepackage{epsfig,bm}

\def\B{$B$}
\def\M{$M$}
\begin{document}

\title[Physics Revealed at Intermediate $p_T$]{Physics Revealed at Intermediate $p_T$}

\author{Rudolph C. Hwa}

\address{Institute of Theoretical Science and Department of Physics, University of Oregon, Eugene, OR 97403-5203, USA}
\ead{hwa@uoregon.edu}
\begin{abstract}
A review is given on the subject of hadron production at intermediate $p_T$ in heavy-ion collisions.  The underlying dynamical processes are inferred from interpreting the data in the framework of recombination.  Ridge formation with or without triggers is found to play an important role in nearly all observables in that $p_T$ region. Correlation data would be hard to interpret without taking ridges into account. The semi-hard partons that create the ridges may even be able to drive elliptic flow without fast thermalization.
\end{abstract}

\section{Introduction}

For heavy-ion collisions at RHIC with $\sqrt{s} = 200$ GeV, the intermediate $p_T$ region usually refers to the interval $2 < p_T < 6$ GeV/c.  At lower $p_T$ hydrodynamical studies have been successful \cite{kh,ph}, while at higher $p_T$ pQCD is more relevant \cite{wg,mg}.  In the intermediate region there is no rigorous theoretical framework that is reliable, but that is where the action is, albeit experimental.  This talk is aimed at the question of what we can learn from the abundant data.

In an overview of the various observables in the intermediate $p_T$ region one can list many topics, starting from single-particle distribution (in $p_T$, $\phi$, and $\eta$) to two-particle correlations (near and away sides), to three-particle correlations (one or two triggers) and autocorrelation (no trigger).  There is not enough time to discuss them all, so I will restrict myself to just the single- and two-particle distributions.  There are many features in $p_T$, $\phi$, and $\eta$, and various nomenclature, such as ridge, head, shoulder, etc.  To tie them all together, there is a need for a theoretical framework to discuss them and to relate them to a common root at the parton level.  That framework, as I shall use, is the recombination model for the hadronization processes at  intermediate $p_T$, leaving open the question about the nature of the partons that hadronize.   Recombination is only a window through which we can ``see'' the partons.  For hadron $p_T<6$ GeV/c those partons are $u$, $d$, and $s$ quarks, the gluons being converted to quarks before hadronization.  The more massive quarks, $c$, $b$, $t$, are primordial and will not enter into our consideration about the medium effect. 

\section[Single-Particle Distributions]{Single-Particle Distributions}

\subsection[Transverse momentum]{Transverse momentum}

The most striking phenomenon at intermediate $p_T$ is the large baryon (\B) to meson (\M) ratio, which peaks at $p_T \approx 3$ GeV/c with $p/\pi \sim 1$ and $\Lambda/K \sim 2$ \cite{sa,ka,ba,ja}.  All three versions of the recombination/coalescence model (ReCo) have been able to reproduce the data on $B/M$ ratio \cite{gr,fr,hy,hy2}.  In the past the unexpected high rate of production of proton has been referred to as  baryon anomaly, which implies that fragmentation is normal.  Not anymore.  On the contrary, high $B/M$ ratio is a signature of ReCo.  It is not dependent on the flow characteristics of the hadrons, since in dAu collision there is no flow, yet the Cronin effect for proton production is higher than that of pion \cite{jc}.  That phenomenon can be explained in the recombination model (RM) \cite{hy3}.  The basic reason is simply that in forming a \B\ or \M\ at the same $p_T$, \B\ needs less quark momenta than \M, and the distribution of quarks decreases rapidly with increasing transverse momentum.

\subsection[Elliptic flow]{Elliptic flow}

At low $p_T$ the recombination of thermal partons is the dominant process, so for \M\ (\B) it is the TT (TTT) component that is more important.  The elliptic flow coefficient $v _2$ is thus related to those of the quarks by:  $v _2^M (p_T) = \sum^2_{i=1}v_2^T (q_i)$ and $v _2^B (p_T) = \sum^3_{i=1}v_2^T (q_i)$.  If one takes $q_i = p_T/2$ for \M\ and $p_T/3$ for \B, then one obtains $v _2^M (p_T/2)/2 = v _2^B (p_T/3)/3$,  which is referred to as quark number scaling (QNS) \cite{v2}, and is a consequence of naive recombination.  There is good experimental support for QNS, especially when plotted in terms of the transverse kinetic-energy $E_T$ \cite{ja2,AA}.

At larger $E_T$ thermal-shower recombination becomes more important.  In that case the quark momenta are not the same in the TS and TTS components, and $v _2^T (q_1) \neq v_2^S (q_2)$ even if $q_1 = q_2$.  Thus, QNS cannot be expected to hold valid at intermediate $p_T$.  In terms of fragmentation, the breaking of QNS was known many years ago \cite{fr}.  More recently, this has been studied in terms of shower contribution \cite{hy4}.  There seems to be some evidence for QNS breaking in the recent data \cite{ja2}, although more accuracy is needed for $E_T/n _q > 1$ GeV.

\subsection[Forward production]{Forward production}

BRAHMS has data from AuAu collisions at 62.4 GeV where both $\eta$ and $p_T$ are measured with $\eta \approx 3.2$ and $p_T$ up to 2.5 GeV/c \cite{ia}.  One can then deduce that many of the data points are very near the kinematic boundary $x_F = 1.0$.  Since shower partons are highly suppressed at high $x$, only thermal partons are important for hadronization at $1 < p_T < 2.5$ GeV/c.  Moreover, antiquarks are not abundant at large $\eta$, so only thermal quarks can contribute to the particles measured  at $\eta \approx 3.2$.  The logical conclusion is that the hadrons detected are mainly protons formed by TTT recombination of light quarks.   The prediction is then that $p/\pi$ is large in forward production \cite{hy5}.

Very recently, BRAHMS has obtained preliminary result that shows the $p/\pi^+$ ratio to be as high as 10 for $p_T <1.2$ GeV/c \cite{rd}.  That is higher than what was predicted in Ref. \cite{hy5}.  The source of the disagreement is due to the over-estimate of the $\bar{p}/p$ ratio  taken from the preliminary rough data given in \cite{hyc}, i.e. $\bar{p}/p \sim 0.05$.  In Arsene's talk at QM08 that ratio was given at $\sim 0.02$.  The lowering of $\bar{p}/p$ ratio significantly increases $p/\pi^+$ ratio because antiquarks are involved in both $\bar{p}$ and $\pi^+$.  In a calculation of the $\bar{q}$ distribution in the forward direction, it is necessary to know the degree of degradation of the forward momentum of incident partons and the subsequent regeneration of $q\bar{q}$ pairs from the lost energy.  The updating of that calculation to fit both $\bar{p}/p$ and $p/\pi^+$ ratios simultaneously is currently being pursued.

\section[Two-Particle Correlation]{Two-Particle Correlation}

There is a wealth of data on two-particle correlation.  Rapidity correlation is the oldest, dating back to the 70s in hadronic collisions \cite{kd}.  The most active area in recent years has been the use of triggers at intermediate or high $p_T$ and the observation of associated particles at various values of $\eta$ and $\phi$ relative to the trigger \cite{ja3,mvl,ht}.  Among the new features found, the most stimulating ones are the discovery of ridges on the near side and the double-hump structure on the away side of $\Delta \phi$.  I shall spend most of the time available to discuss the implications of the former.

\subsection[Ridgeology]{Ridgeology}

I shall refer to the phenomenology of ridges as ridgeology.  It was shown by Putschke at QM06 that the distribution of particles associated with a trigger in the range $3 <p^{\rm trig}_T < 4$ GeV/c exhibits a peak at small $\Delta \eta$ and $\Delta \phi$ sitting on top of a ridge that has a wide range in $\Delta \eta$ in excess of $\pm 1.5$ \cite{jp}.  STAR collaboration has used the notation $J$ for the peak and $R$ for the ridge, a practice that has been followed by others, although one must be cautious in the realization that $J$ standing for Jet is only a piece of the jet structure, which must include $R$.  There are many features of ridgeology that should be taken into account, if a partonic basis of the phenomenon is to be constructed.  Let us list those features that are shown in \cite{jp}.

\subsubsection[Centrality dependence]{Centrality dependence}

The yield in $R$ for $p_T^{\rm assoc} > 2$ GeV/c,  integrated over $\Delta \eta$ and $\Delta \phi$, decreases with $N_{\rm part}$ and vanishes as $N_{\rm part}$ approaches the minimum corresponding to $pp$ collisions.  Thus the formation of $R$ depends on the nuclear medium.

\subsubsection[Dependence on $p_T^{\rm trig}$]{Dependence on $p_T^{\rm trig}$}

The ridge yield decreases only slightly for 0-10\% centrality, as $p_T^{\rm trig}$ is increased from 3 to 9 GeV/c.  Thus ridge is strongly correlated to jet production.  Since trigger bias favors the detection of jets produced by hard partons that do not lose much energy by traversing the nuclear medium, ridge is therefore due to the medium effect near the surface.

\subsubsection[Dependence on $p_T^{\rm assoc}$]{Dependence on $p_T^{\rm assoc}$}

The ridge yield is exponential in its dependence on $p_T^{\rm assoc}$; the slope in the semi-log plot is essentially independent of $p_T^{\rm trig}$.  Exponential behavior means that the particles in the ridge are emitted from a thermal source.  Usually thermal partons are regarded as being uncorrelated.  In this case they are all correlated to the semi-hard parton that initiates the jet.  We thus interpret the observed characteristics as indicating that the ridge is from a thermal source enhanced by the energy lost by the semi-hard parton traversing the medium.

\subsubsection[$B/M$ ratio in the ridge]{$B/M$ ratio in the ridge}

The $\Lambda/K^0_s$ ratio in the ridge is found to be around 1 at $p_T^{\rm assoc} \sim 2$ GeV/c \cite{jb}.  The $p/\pi$ ratio reported by Putschke in his talk at QM06 (but not in his write-up \cite{jp}) is the ratio in $R$ relative to that in $J$; it is very large $(>2)$ for $3 < p_T^{\rm trig} < 5$ GeV/c.  Thus in the ridge the $p/\pi$ ratio is consistent with the $\Lambda/K^0_s$ ratio.  That can be understood only in the framework in which the ridge hadrons are formed by recombination of enhanced thermal partons.

Taking into account all these properties described in the four subsections above, it would be hard to construct a model that can differ significantly from the one we outline below.  There are several stages of the dynamical process:  (a) a semi-hard scattering occurs near the surface, (b) as one of the semi-hard parton traverses the medium on its way out, it loses energy, (c) the energy lost to the medium enhances the thermal partons in the vicinity of the trajectory, (d) the semi-hard parton that emerges from the surface generates shower partons S, (e) the recombination of S with the enhanced thermal partons T forms hadrons that can be either the trigger or an associated particle in $J$, (f) the recombination of TT or TTT forms $M$ or $B$ in $R$, (g) particles in $J$ involve S, so they stay close to the trigger direction, but enhanced thermal partons in $T$ can flow with medium expansion longitudinally and acquire large $\Delta \eta$, but not azimuthally, thus staying restricted in $\Delta \phi$, and $(h)$ the enhancement of T over the bulk results in the ridge after background subtraction.  None of these subprocesses can be calculated reliably either in pQCD or in hydrodynamics, but the physical reasoning behind each one of them is constrained by the data observed.  A model to quantify the processes was advanced in \cite{ch}, when very preliminary data were available \cite{ja3}.  Now, with more abundant data at hand, more aspects of the model should be pinned down with less uncertainty.  Moreover, some light should be shed on the data from PHENIX whose lower  $\eta$ acceptance hinders the study of ridges.

\subsection[Consequences of ridgeology]{Consequences of ridgeology}

With the framework of ridge formation described above in mind, we can now revisit some of the basic observables, in particular, single-particle distribution, correlations and elliptic flow.

\subsubsection[Effect of ridges on single-particle spectra]{Effect of ridges on single-particle spectra}

It is important to note that although the foregoing study of ridgeology is based on events with triggers, ridges are present with or without triggers.  That is because the ridges are induced by semi-hard scattering which can take place whether or not a hadron in a chosen $p_T$ range is used to select events.  Experimentally, it is known that the peak and ridge structure is seen in auto-correlation where no triggers are used \cite{star2}.  The implication of that is that the ridge hadrons are pervasive and are always present in the single-particle spectra.

In the RM the pion spectra can be understood in terms of a combination of TT, TS and SS contributions.  The last term, SS recombination, is equivalent to fragmentation, since that is the basis on which the shower partons are determined \cite{hy6}.  In the intermediate $p_T$ region the TS term dominates, while at low $p_T$ TT is most important.  Since semi-hard partons generate the ridges, whose hadron $p_T$ can extend into the intermediate range, the thermal partons in T in TS and TT include the enhanced ones, not just the bulk.  To find evidence for that, we would like to isolate the TT component even at intermediate $p_T$.  That may be hard experimentally, but we can consider a circuitous way.

Since the production of $s$ quark in the shower is suppressed, the dominant mechanism for the production of $\Omega$ in the intermediate $p_T$ region is by TTT recombination of $s$ quarks.  If so, then the $p_T$ spectrum of $\Omega$ should be exponential \cite{hy7}.  Indeed, there is experimental evidence for that for $p_T$ up to 5.5 GeV/c \cite{ja4,sb}.  To find exponential behavior up to such a high $p_T$ is remarkable, but is not sufficient to conclude that it contains ridge particles.  The latter requires correlation study.

At QM06  Bielcikova showed that using $\Omega$ as a trigger particle there are associated hadrons \cite{jb3}.  That seems puzzling at first \cite{rh}, since thermal hadrons do not normally have correlated partners.  The resolution of that puzzle is in the recognition that both the $\Omega$ itself as the trigger and its associated particles are in the ridge, whose underlying partons are thermal, hence exponential, but are correlated to the initiating semi-hard parton.  These properties have been demonstrated to be consistent with data in the RM \cite{ch2}.  Since the peak in $\Delta \phi$ arises entirely from the ridge, the predication is that there would be no peak ($J$) in the $\Delta \eta$ distribution.

\subsubsection[Jet correlation]{Jet correlation}

The correlation characteristics of trigger and partner in jets have been studied by PHENIX \cite{AA2}, as well as by STAR \cite{jp,jb}.  Since the $\eta$-acceptance of the PHENIX detector is for $|\eta|<0.35$ only, it is difficult to isolate the ridge contribution from the peak ($J$) that sits on top of the ridge.  Not seeing the ridge does not mean that it is not present in both $\Delta \eta$ and $\Delta \phi$ distributions.  Focusing on only the $\Delta \phi$ distribution where $R$ and $J$ are merged, any interpretation in terms of fragmentation of jets can be misleading, especially when the trigger momentum is not high:  $2.5 < p_T^{\rm trig} < 4$ GeV/c \cite{AA2}.  Correlation in $R$ is different from that in $J$, both being different from jet fragmentation which corresponds to SS recombination at $p_T > 6$ GeV/c.

In \cite{AA2} is shown yield/trigger for meson-meson correlation increasing monotonically with $N_{\rm part}$.  In \cite{jb4} STAR shows the same for $h$-$h$ correlation when $J$ and $R$ are combined, but for $J$ by itself the yield is constant in $N_{\rm part}$.  Thus the increasing part of $J + R$ is due entirely to $R$, which is not included in the interpretation of the PHENIX data.  The constancy of $J$ in $N_{\rm part}$ can be understood as the combination of TS and SS components, the former increasing with $N_{\rm part}$ because of the medium contribution to the thermal partons, while the latter decreasing with $N_{\rm part}$, since the nuclear medium degrades the semi-hard parton momentum.  At very low $N_{\rm part}$ corresponding to $pp$ collision, there is only jet fragmentation, so SS dominates even for $3 < p_T^{\rm trig} < 5$ GeV/c in the STAR analysis.  The rising $R$ contribution to what is regarded as jet yield in \cite{AA2} is due to TT recombination, which is not uncorrelated to the trigger, since it would not be there without semi-had scattering.

The $B/M$ ratio of the associated particles increases with $p_T^{\rm assoc}$ to about 0.2 at $p_T^{\rm assoc} \approx 1.8$ GeV/c for $2.5 < p^{\rm trig}_T < 4.0$ GeV/c \cite{saf}.  That ratio will continue to rise at higher $p^{\rm assoc}_T$, the main contributor to that rise will come from the $R$ component.  That is not a speculation, but can be inferred from the STAR data where the relative ridge yields (i.e. $R/J$) for $p$ and $\pi$ are roughly $4.8 \pm 0.8$ and $2.2 \pm 0.3$, respectively, at $p^{\rm trig}_T \sim 3.5$ GeV/c and $p^{\rm assoc}_T > 2$ GeV/c \cite{jp}.  Thus any correlation among particles in a jet at intermediate $p_T$ cannot be properly understood without taking ridge into account.

\subsubsection[Effect of ridges on elliptic flow]{Effect of ridges on elliptic flow}

The conventional approach to elliptic flow is by use of hydrodynamics which produces satisfactory results for $p_T < 1.5$ GeV/c \cite{kh,ph}.  However, it requires fast thermalization, i.e., $\tau_0 = 0.6$ fm/c, the validity of which has never been substantiated in QCD.  What if $\tau_0$ can never be less than 1 fm/c?  How much of the perfect fluid picture has to be given up?  How can we understand $v_2$ without high pressure gradient at early time?

Ridges offer an alternative way to understand elliptic flow at low $p_T$.  If semi-hard jets are soft enough, there are many of them:  if $q_T \sim 2$-$3$ GeV/c, then $x \sim 0.03$ at which the density of soft partons is high.  Yet they are hard enough so that the time scale involved is $ \sim q^{-1}_T \sim 0.1$ fm/c, which is earlier than any thermalization time contemplated.  In a non-central collision at impact parameter $b$ the boundary of the almond-shaped overlap region in the transverse plane has a maximum opening angle of $\Phi = \cos^{-1}(b/2R_A)$.  A semi-hard scattering near the surface at any $|\phi| < \Phi$ sends a jet, on average, in the direction normal to the surface, that being the only angle in the geometry of the problem.  Since there are many such jets in each AA collision, there is a layer of ridges at the surface without triggers.  That is what drives the elliptic flow \cite{rh2}.

It is possible to show by simple geometrical consideration that
\begin{eqnarray}
v_2 (p_T, b) = {\sin 2 \Phi (b) \over \pi B (p_T)/R(p_T) + 2 \Phi (b)} ,
\label{1}
\end{eqnarray}
where $B (p_T)$ and $R(p_T)$ are the $p_T$ distributions of the hadrons in the bulk and ridge, respectively.  Since they cannot be calculated, we use the data as inputs for them in (\ref{1}), thereby relating ridgeology to $v _2$.  The agreement with data for all centralities for both $\pi$ and $p$ is very good \cite{hy4,  rh2}.  For $p_T$ in the intermediate region the contribution from the shower partons must be included.  As noted earlier, the quark number scaling is broken.

\subsection[Away-side structure]{Away-side structure}

There is interesting structure in the $\Delta \phi$ distributions on the away side; it has stimulated a great deal of interest both experimentally and theoretically \cite{large}.  Neither time (at QM08) nor space (here) allows for extensive discussion on the subject.  Let it be mentioned that the double-hump structure on the two sides of $\Delta \phi = \pi$, observed first by STAR \cite{ja5} and then by PHENIX \cite{AA3} that shows only mild dependence of the distance between the two peaks on $p_T^{\rm assoc}$, seems to favor Mach cone and deflected jets, as opposed to gluon radiation.  Hadrons in the humps have exponential behavior in $p_T^{\rm assoc}$ and have large $B/M$ ratio; hence, they strongly suggest a possible relationship between the humps on the away side and the ridges on the near side.  While for deflected jets that can be understood in terms of TT and TS recombination \cite{ch3}, it would be harder to describe the Mach cone properties at the parton level.

\section[On to LHC]{On to LHC}

Many predictions have been made on what to expect at LHC \cite{ar}.  Those with existing codes can make extrapolations to higher energy and show, for example, where the yield from pQCD calculation dominates over the hydro result.  My interest is rather to ask the question whether there is any new physics that cannot be obtained by extrapolation.  Since the density of semi-hard partons is so high at LHC, their close proximity to one another in each event creates a new possibility not considered at lower energies.  Shower partons from near-by semi-hard partons can recombine to form $\pi$ or $p$.  The $p/ \pi$ ratio of such hadrons would then be very large even for $10 < p_T < 20$ GeV/c, since $p$ requires less parton momenta than $\pi$.  Furthermore, for any such high $p_T$ hadrons used as trigger there would be no associated particles distinguishable from the background, which will consist of many similar particles produced by SS and SSS recombination due to the abundance of semi-hard partons \cite{hy8}.  

Since semi-hard scattering is not accounted for by hydro, the particles produced by the mechanism above are not a part of the hydro flow, yet they are uncorrelated and belong to the background.  Thus at LHC we expect a mismatch between hydro and background.  Indeed, it is not clear whether ridges can be identified.  The physics in the intermediate $p_T$ region at LHC is likely to be very different from that at RHIC.

\ack

I am grateful to M.\ Tannenbaum for helpful communication.  This work was supported,  in part,  by the U.\ S.\ Department of Energy under Grant No. DE-FG02-96ER40972.


\section*{References}
\begin{thebibliography}{10}

\bibitem{kh} 
P.\ F.\ Koch and V.\ Heinz,  in {\it QGP3}, edited by R.\ C.\ Hwa and X.\ N.\ Wang, World Scientific, 2004.

\bibitem{ph} 
P.\ Houvinen,  QGP3in {\it QGP3}, edited by R.\ C.\ Hwa and X.\ N.\ Wang, World Scientific, 2004.

\bibitem{wg} 
X.\ N.\ Wang and  M.\  Gyulassy, Phys.\ Rev.\ Lett.\ {\bf 68}, 1480 (1992); Phys.\ Rev.\ Lett.\ {\bf 86}, 3496 (2001).

\bibitem{mg} 
M.\  Gyulassy {\it et al.}, in {\it QGP3}, edited by R.\ C.\ Hwa and X.\ N.\ Wang, World Scientific, 2004.

\bibitem{sa} 
S.\ S.\ Adler {\it et al.} (PHENIX), Phys.\ Rev.\ Lett.\ {\bf 91}, 172301 (2003). 

\bibitem{ka}
K.\ Adcox  {\it et al.} (PHENIX), Nucl.\ Phys.\  A{\bf 757}, 184 (2005).

\bibitem{ba}
B.\  I.\ Abelev {\it et al.}, (STAR), Phys.\ Rev.\ Lett.\ {\bf 97}, 152301 (2006).

\bibitem{ja} 
J.\ Adams {\it et al.} (STAR), nucl-ex/0601042.

\bibitem{gr} 
V.\ Greco, C.\ M.\ Ko, and P.\ L\'{e}vai, Phys.\ Rev.\
Lett.\ {\bf 90}, 202302 (2003); Phys.\ Rev.\ C {\bf 68}, 034904 (2003).

\bibitem{fr} 
R.\ J.\ Fries, B. M\"{u}ller, C.\ Nonaka and S.\ A.\
Bass, Phys.\ Rev.\ Lett.\ {\bf 90}, 202303 (2003);  Phys.\ Rev.\ C
{\bf 68}, 044902 (2003).

\bibitem{hy} R.\ C.\ Hwa, and C.\ B.\ Yang, Phys.\ Rev.\ C {\bf 67},
034902 (2003).

\bibitem{hy2}
R.\ C.\ Hwa and C.\ B.\  Yang,  Phys.\ Rev.\ C {\bf 70}, 024905 (2004).

\bibitem{jc} J.\ W.\ Cronin {\it et al.}, Phys.\ Rev.\ D {\bf 11}, 3105 (1975); D.\ Antreasyan {\it et al.}, Phys.\ Rev.\ D {\bf 19}, 764 (1979).

\bibitem{hy3}
R.\ C.\ Hwa and C.\ B.\  Yang, Phys.\ Rev.\ Lett.\ {\bf 93}, 082302 (2004); Phys.\ Rev.\ C {\bf 70}, 037901(2004).

\bibitem{v2} 
D.\ Moln{\' a}r and S.\ A.\ Voloshin,  Phys.\ Rev.\
Lett.\ {\bf 91}, 092301 (2003).

\bibitem{ja2}
J.\ Adams {\it et al.}, (STAR), Phys.\ Rev.\ C  {\bf 72}, 014904 (2005); B.\  I.\ Abelev {\it et al.}, (STAR), Phys.\ Rev.\ C  {\bf 75}, 054906 (2007).

\bibitem{AA} A.\ Adare {\it et al.} (PHENIX), Phys.\ Rev.\ Lett.\ {\bf 98}, 162301 (2007).

\bibitem{hy4} R.\ C.\ Hwa, and C.\ B.\ Yang, arXiv:  0801.2183.

\bibitem{ia}I.\ Arsene {\it et al.}\ (BRAHMS),
nucl-ex/0602018.

\bibitem{hy5}R.\ C.\ Hwa and C.\ B.\  Yang, Phys.\ Rev.\ C {\bf 76}, 014901 (2007).

\bibitem{rd} 
R.\ Debbe, plenary talk at Quark Matter 2008, Jaipur, India; P.\ Staszel, private communication.

\bibitem{hyc}
H.\ Yang (BRAHMS), Czech J.\ Phys.\ {\bf 56}, A27 (2006). 

\bibitem{kd}
W.\ Kittel and E.\ A.\ De Wolf, {\it Soft Multihadron Dynamics}, (World Scientific, Singapore, 2005). 

\bibitem{ja3} 
J.\ Adams {\it et al.} (STAR), Phys.\ Rev.\ Lett.\ {\bf 95}, 152301 (2005).

\bibitem{mvl}
M.\ van Leeuwen, J.\ Phys.\ G:  Nucl.\ Part.\ Phys.\ {\bf 34}, S559 (2007). 

\bibitem{ht} R.\ C. Hwa, Nucl.\ Phys.\  {\bf A783}, 57c (2007);  Int.\ J.\ Mod.\  Phys.\  {\bf E16} 3176 (2007).

\bibitem{jp} J.\ Putschke (for STAR), J.\ Phys.\ G:  Nucl.\ Part.\ Phys.\ {\bf 34}, S679 (2007), talk given at QM06.

\bibitem{jb} 
J.\ Bielcikova (for STAR),  arXiv: 0707.3100.

\bibitem{ch} 
C.\ B.\ Chiu and R.\ C.\ Hwa, Phys.\ Rev.\ C {\bf 72}, 034903 (2005).

\bibitem{star2} J.\ Adams {\it et al.}, (STAR) Phys.\ Rev.\ C {\bf 73}, 064907 (2006).

\bibitem{hy6} 
R.\ C.\ Hwa and C.\ B.\ Yang, Phys.\ Rev.\ C  {\bf 70}, 024904 (2004); Phys.\ Rev.\ C {\bf 73}, 064904
(2006).

\bibitem{hy7}
R.\ C.\ Hwa and C.\ B.\ Yang, Phys.\ Rev.\ C {\bf 75}, 054904 (2007).

\bibitem{ja4} 
J.\ Adams {\it et al.} (STAR), Phys.\ Rev.\ Lett.\ {\bf 98}, 062301 (2007).

\bibitem{sb} 
S.\ L.\ Blyth (for STAR), J.\ Phys.\ G:  Nucl.\ Part.\ Phys.\ {\bf 34}, S933 (2007), talk given at QM06.

\bibitem{jb3} 
J.\ Bielcikova (for STAR), J.\ Phys.\ G:  Nucl.\ Part.\ Phys.\ {\bf 34}, S929 (2007), talk given at QM06.

\bibitem{rh} 
R.\ C.\ Hwa, J.\ Phys.\ G:  Nucl.\ Part.\ Phys.\ {\bf 34}, S789 (2007).

\bibitem{ch2} 
C.\ B.\ Chiu and R.\ C.\ Hwa, Phys.\ Rev.\ C {\bf 76}, 024904 (2007).

\bibitem{AA2}  
A.\ Adare {\it et al.} (PHENIX), Phys.\  Lett.\ B{\bf 649}, 359 (2007).

\bibitem{jb4} 
J.\ Bielcikova (for STAR), talk presented at Hard Probes  2006, Nucl.\ Phys.\ A{\bf 783}, 565c (2007).

\bibitem{saf} 
S.\ Afanasiev {\it et al.}  (STAR),  arXiv: 0712.3033.

\bibitem{rh2} 
R.\ C.\ Hwa,  arXiv: 0708.1508.

\bibitem{large} 
See the large numbers of papers presented at the plenary and parallel sessions at QM08.

\bibitem{ja5} 
J.\ Adams {\it et al.} (STAR), Phys.\ Rev.\ Lett.\ {\bf 95}, 152301 (2005).

\bibitem{AA3} 
A.\ Adare {\it et al.}, (PHENIX), arXiv: 0705.3238.

\bibitem{ch3} 
C.\ B.\ Chiu and R.\ C.\ Hwa, Phys.\ Rev.\ C {\bf 74}, 064906 (2006).

\bibitem{ar} 
N.\ Armesto,  arXiv: 0711.0974.

\bibitem{hy8} 
R.\ C.\ Hwa and C.\ B.\ Yang, Phys.\ Rev.\ Lett.\ {\bf 97}, 042301 (2006).

\end{thebibliography}
\end{document}